\documentclass[prl,twocolumn]{revtex4}
\usepackage{graphicx}

\newcommand{\sSC}{{\scriptscriptstyle \text{SC}}}
\newcommand{\sMF}{{\scriptscriptstyle \text{MF}}}
\newcommand{\sQ}{{\scriptscriptstyle Q}}
\newcommand{\sB}{{\scriptscriptstyle B}}
\newcommand{\sLambda}{{\scriptscriptstyle \Lambda}}

\begin{document}

\title{Gaussian superconducting fluctuations, thermal
transport, and the Nernst effect}

\author{Iddo Ussishkin}
\author{S. L. Sondhi}
\author{David A. Huse}
\affiliation{Department of Physics, Princeton University, Princeton, NJ
08544}

\date{April 22, 2002}

\begin{abstract}
We calculate the contribution of superconducting fluctuations to thermal
transport in the normal state, at low magnetic fields. We do so in the
Gaussian approximation to their critical dynamics which is also the
Aslamazov-Larkin approximation in the microscopics. Our results for the
thermal conductivity tensor and the transverse thermoelectric response are
new. The latter compare favorably with the data of Ong and collaborators on
the Nernst effect in the cuprates.
\end{abstract}

\maketitle

The study of fluctuations in superconductors~\cite{reviews} had a marked
revival after the discovery of the cuprate superconductors. It was realized
early that their short coherence lengths produce a large regime of strong
fluctuation phenomena~\cite{Fisher-Fisher-Huse}. More recently the
``pseudogap'' region of the cuprate phase diagram has come into sharp
focus, and one line of thought attributes its features to strong
superconducting fluctuations~\cite{Emery-Kivelson,Randeria}.  Among the
most striking experimental findings that plausibly support this link are
recent measurements by Ong's group~\cite{Ong-etal} where the Nernst effect
becomes sizeable far above $T_c$ in the underdoped regime of the
high-temperature superconductors. As the Nernst effect is small in ordinary
metals and large in the vortex state of superconductors the {\it a priori}
case for crediting superconducting fluctuations is strong.

In this Letter we calculate the contribution of Gaussian
superconducting fluctuations to the transverse thermoelectric response
above $T_c$ in the low magnetic field limit---the simplest computation
that can be used for a quantitative comparison with these experimental
results. This is already a non-trivial exercise as it requires a proper
subtraction of {\it bulk} magnetization currents that enter naive
expressions~\cite{Cooper-Halperin-Ruzin}. We also give results for the
thermal conductivity tensor, thus presenting a full picture for
Gaussian thermal transport in the system.

Our result for the transverse thermoelectric response depends only on the
superconducting coherence length $\xi$, making it particularly suitable for
comparison with experiment. We present a comparison with data from
different samples of La$_{2-x}$Sr$_x$CuO$_4$ (LSCO). We find that the
Gaussian Nernst effect, using the actual $T_c$, is quantitatively
comparable to the measured signal in the optimally doped and overdoped
samples. For the underdoped sample the measured signal in the ``pseudogap''
region is larger and requires postulating a suppression of the actual $T_c$
from the mean-field $T_c$, consistent with the superconducting fluctuations
scenario, in order to achieve an understanding of its magnitude, as we
illustrate by a self-consistent Hartree computation.

Before proceeding, we note that the traditional description of the Nernst
effect is of thermally driven vortices producing a transverse voltage via
phase slips. In a fluctuation regime this description is only convenient if
the vortices provide an effective parameterization, as we expect will be
the case near a two dimensional Kosterlitz-Thouless transition. At higher
temperatures, where the vortices and anti-vortices are strongly
overlapping, a different simplification involving Gaussian fluctuations
becomes available and we will take recourse to that in the present paper.

\emph{Gaussian computations.}---Our calculations can be carried out in
two equivalent ways. First, they involve following Aslamazov and
Larkin~\cite{Aslamazov-Larkin} and keeping the Feynman diagrams that
now bear their names and arise in a Gaussian treatment of the quantum
functional integral. Second, as was noted, e.g., in the case of
paraconductivity~\cite{reviews}, one can keep Gaussian fluctuations in
a stochastic time-dependent Ginzburg-Landau equation (TDGL) that is
designed to recover the equilibrium Ginzburg-Landau free energy. At
this level, with the coefficients of the TDGL derived from BCS
microscopics near the mean-field $T_c$, the two computations are
identical. The TDGL, however, has a second interpretation---as a model
of the critical dynamics traditionally assumed for superconductors
(model A)~\cite{Hohenberg-Halperin} which should have a wider validity,
with the coefficients no longer constrained by BCS microscopics. In the
following we will mostly use the TDGL description, while appealing to
the microscopics to justify the form of the current
operators~\cite{Vishveshwara-Fisher}.

The stochastic TDGL is defined by the Ginzburg-Landau free energy
\begin{equation}\label{GL-free-energy}
\mathcal{F} = \int d \mathbf{x} \left[ a |\psi|^2 + \frac{b}{2} |\psi|^4 +
\frac{\hbar^2}{2 m^*} \left| \left( \nabla - i \frac{e^*}{\hbar c}
\mathbf{A} \right) \psi \right|^2 \right] ,
\end{equation}
where $a = a_0 (T - T_c)$, and the time evolution
\begin{equation}\label{TDGL}
(\tau + i \tau') \left( \frac{\partial}{\partial t} + i \frac{e^*}{\hbar}
\phi \right) \psi = - \frac{\delta \mathcal{F}}{\delta \psi^*} + \zeta .
\end{equation}
Equation~(\ref{TDGL}) describes the decay of the order parameter
configuration towards its minimal free energy, interrupted by thermal
fluctuations introduced through the white noise $\zeta$ with correlator
\begin{equation}\label{noise-correlator}
\langle \zeta^* (\mathbf {r}, t) \zeta (\mathbf {r'}, t') \rangle = 2 T
\tau \, \delta (\mathbf{r} - \mathbf{r'}) \delta (t - t') .
\end{equation}
The relaxation time of the order parameter is $\tau$, and $\tau'$ leads to
the precession of the order parameter during the relaxation process (in
this convention, $\tau / \hbar$ and $\tau' / \hbar$ are dimensionless). In
the presence of particle-hole symmetry $\tau' = 0$.

To calculate the response of the system, Eq.~(\ref{TDGL}) is to be
solved in the presence of the appropriate driving field (electric field
or temperature gradient). Note that a temperature gradient appears in
the TDGL both in the parameter $a$ and in the noise correlator,
Eq.~(\ref{noise-correlator}). At the level of Gaussian fluctuations the
TDGL is a linear equation, with a solution of the the form $\psi
(\mathbf{r}, t) = \int d \mathbf{r'} \, d t ' \, G (\mathbf{r}, t;
\mathbf{r'}, t') \zeta (\mathbf{r'}, t')$, where $G$ is the Green
function of the TDGL. This expression may then be used to calculate the
electric and heat currents,
\begin{equation}\label{electric-current}
\mathbf{j}^e = -i \frac{e^* \hbar}{2 m^*} \left\langle \psi^* \left(
\nabla - i \frac{e^*}{\hbar c} \mathbf{A} \right) \psi \right\rangle +
\text{c.c.},
\end{equation}
\begin{equation}\label{heat-current}
\mathbf{j}^\sQ \! = \! - \frac{\hbar^2}{2 m^*} \left\langle \! \left(
\frac{\partial}{\partial t} - i \frac{e^*}{\hbar} \phi \right) \! \psi^* \!
\left( \nabla - \frac{i e^*}{\hbar c} \mathbf{A} \right) \! \psi \!
\right\rangle + \text{c.c.}
\end{equation}
Ullah and Dorsey~\cite{Ullah-Dorsey} used precisely the same model.

\emph{Currents and fields.}---As model A has no conservation laws, the
traditional route to identifying currents is not available. While it is
possible to construct hydrodynamic arguments that justify the forms in
(\ref{electric-current}) and (\ref{heat-current})~\cite{Schmid}, for our
purposes it is sufficient to appeal to the microscopics of the
Aslamazov-Larkin contribution, and show how the heat current arises from
the appropriate vertex in the microscopic theory.~\cite{Caroli-Maki}

The microscopic theory is perhaps best recast in a functional integral
approach. By means of the Hubbard-Stratonovich decoupling of the
interaction, the expectation value of a current operator is expressed
as an imaginary time functional integral over the pairing field,
\begin{equation}\label{operator-average}
\langle \, \mathbf{j} \, \rangle = \frac{ \int D \psi \, D \bar \psi \,
\langle \, \mathbf{j} \, \rangle_{\psi \bar \psi \phi} \, e^{
- S_{\text{eff}} (\psi, \bar \psi, \phi)} }{ \int D \psi \, D \bar \psi \,
e^{-S_{\text{eff}} (\psi, \bar \psi, \phi )} },
\end{equation}
where $S_{\text{eff}} (\psi, \bar \psi, \phi)$ is the effective action
for the pairing field $\psi$, in presence of a potential $\phi$, and
$\langle \, \mathbf{j} \, \rangle_{\psi \bar \psi \phi}$ is the current
of a non-interacting Fermi gas driven by electric and pairing fields.
Ignoring the $\phi$ dependence of $\langle \, \mathbf{j} \, \rangle$ is
the Aslamazov-Larkin approximation; the expectation value is the vertex
in their diagram and its long wavelength, low frequency form is the
TDGL current. For the electric current this was done in the original
calculation~\cite{Aslamazov-Larkin}. Using the microscopic heat current
operator, one obtains a relation between heat current and electric
current vertices, $\mathbf{J}^\sQ = - \omega \mathbf{J}^e / 2 e$,
independent of disorder. This corrects a factor of 2 in the calculation
of Reizer and Sergeev~\cite{Reizer-Sergeev}. The heat current in the
TDGL, Eq.~(\ref{heat-current}), is now immediate by this result.

To obtain a response to a temperature gradient in the microscopic theory,
one calculates the response to a ``gravitational field'', coupled to the
energy density in the Hamiltonian, and then uses an Einstein relation to
obtain the response to a temperature gradient~\cite{Luttinger}. The
microscopic calculation of the Aslamazov-Larkin diagrams leads to the same
results obtained using the TDGL approach,
Eqs.~(\ref{GL-free-energy})--(\ref{heat-current}), where the response to a
temperature gradient is obtained directly. Ignored here are the
Maki-Thompson and density of state corrections to the normal state
response~\cite{reviews,Abrahams-Redi-Woo}.

\emph{The transport coefficients} are defined by the standard linear
response relations,
\begin{equation}\label{response-coefficients}
  \left( \begin{array}{c} \mathbf{j}_{\text{tr}}^e \\
  \mathbf{j}_{\text{tr}}^\sQ \end{array} \right) =
  \left( \begin{array}{ccc} \sigma & \alpha  \\ \tilde \alpha &
  \kappa \end{array} \right)
  \left( \begin{array}{c} \mathbf{E} \\ - \nabla T \end{array} \right) \, .
\end{equation}
The off-diagonal thermoelectric tensors obey the Onsager relations $\tilde
\alpha = T \alpha$. Below, we calculate the contribution of Gaussian
superconducting fluctuations to the thermoelectric and the thermal
conductivity tensors ($\alpha^\sSC$ and $\kappa^\sSC$). We do not consider
here the electrical conductivity tensor $\sigma^\sSC$, which has been
studied extensively~\cite{reviews}.

Before proceeding, we note that particle-hole symmetry ($\tau'=0$) implies
that $\sigma_{xy}^\sSC = \alpha_{xx}^\sSC = \kappa_{xy}^\sSC = 0$. In the
following we break particle-hole symmetry only when considering these
coefficients. In addition, we calculate the longitudinal coefficients in
the absence of magnetic field $B$, and the transverse coefficients to
linear order in $B$ (valid for $B \xi^2 \ll h c / e^*$). The final results
in both two and three dimensions are presented in
Table~\ref{table:results}. The generalization of these results to a layered
superconductor using a Lawrence-Doniach model is straightforward.

\begin{table}
\caption{Contribution of Gaussian fluctuations to thermoelectric and
thermal transport coefficients. Here, $\ell_\sB = (\hbar c / e B)^{1/2}$ is
the magnetic length, $\xi = \hbar / (2 m^* a)^{1/2}$ is the Ginzburg-Landau
coherence length, $T_\sLambda$ is a cutoff temperature, and $e^* = -2 e$ is
used. Exact coefficients are given for diverging terms only (for
non-diverging contributions, singular behavior is indicated without
specifying prefactors).} \label{table:results}
\begin{ruledtabular}
\begin{tabular}{lcc}
& two dimensions & three dimensions \\
\hline \vspace{-0.3cm} \\
$\alpha_{xx}^\sSC$\footnote{Previously considered in
Refs.~\cite{Ullah-Dorsey,Reizer-Sergeev,Maki-74}.} & $\displaystyle
\frac{1}{2 \pi} \, \frac{e}{\hbar} \, \frac{\tau'}{\tau} \ln \left(
\frac{T_\sLambda}{T - T_c} \right)$ & $c_0 - c_1 \sqrt{T - T_c}$ \\
\vspace{-0.2cm} \\
$\alpha_{xy}^\sSC$ & $\displaystyle \frac{1}{6 \pi} \, \frac{e}{\hbar} \,
\frac{\xi^2}{\ell_\sB^2} \propto \frac {1}{T - T_c}$  & $\displaystyle
\frac{1}{12 \pi} \, \frac{e}{\hbar} \,
\frac{\xi}{\ell_\sB^2} \propto \frac {1}{\sqrt{T - T_c}}$ \\
\vspace{-0.2cm} \\
$\kappa_{xx}^\sSC$ & $\displaystyle c_0 - c_1 (T -
T_c) \ln \left( \frac{T_\sLambda}{T - T_c} \right)$ &
$c_0 - c_1 (T - T_c)^{3/2}$ \\ \vspace{-0.2cm} \\
$\kappa_{xy}^\sSC$ & $\displaystyle \frac{1}{4 \pi} \, \frac{e B}{m^* c} \,
\frac{\hbar \tau'}{\tau^2} \ln \left( \frac{T_\sLambda}{T - T_c} \right)$ &
$c_0 - c_1 \sqrt{T - T_c}$ \\ \vspace{-0.3cm} \\
\end{tabular}
\end{ruledtabular}
\end{table}

The thermoelectric response $\alpha_{xx}^\sSC$ may be calculated in two
ways, either as the heat current response to an electric field or as the
electric current response to a temperature gradient. In this way Onsager
relations are verified. The result has a logarithmic divergence at $T_c$ in
two dimensions~\cite{Ullah-Dorsey,Reizer-Sergeev,Maki-74}.

The calculation of the transverse thermoelectric coefficient
$\alpha_{xy}^\sSC$ raises the issue of equilibrium magnetization currents.
To illustrate this point, consider the result of calculating both the heat
current response to an electric field and the electric current response to
a temperature gradient (in two dimensions),
\begin{equation}\label{jQy-over-Ex}
\frac{j_y^\sQ}{E_x} = - \frac{1}{16 \pi} \, \frac{e^{*2} B T}{m^* c a_0} \,
\frac{1}{T - T_c} \, ,
\end{equation}
\begin{equation}\label{jy-over-nablaTx}
\frac{j_y^e}{(- \nabla T)_x} = \frac{1}{48 \pi} \, \frac{e^{*2} B}{m^* c
a_0} \left( \frac{2 T}{(T - T_c)^2} - \frac{3}{T - T_c} \right) \! .
\end{equation}
Equation~(\ref{jQy-over-Ex}) agrees with the calculations of Ullah and
Dorsey~\cite{Ullah-Dorsey}. However, the two results,
Eqs.~(\ref{jQy-over-Ex})--(\ref{jy-over-nablaTx}), give a different
answer for the transverse thermoelectric response, and are clearly
incompatible with Onsager relations. The reason for this apparent
discrepancy is that Eqs.~(\ref{electric-current})--(\ref{heat-current})
give the \emph{total} currents in the system, containing contributions
from both \emph{transport} and \emph{magnetization} currents. In the
context of superconductivity this issue was raised in
Refs.~\cite{Maki,Hu}, but was later ignored in works regarding
fluctuation
contributions~\cite{Ullah-Dorsey,Varlamov-Livanov-91,Maki-91}. Here we
follow Cooper, Halperin, and Ruzin~\cite{Cooper-Halperin-Ruzin} who
considered the general problem of magnetization currents in
magneto-thermoelectric transport in detail (they encountered it in the
context of the quantum Hall effect).

In the presence of a magnetic field, the system has magnetization currents
in equilibrium. These currents are divergence-free, and as a consequence do
not make any contribution to the net current flows that are measured in a
transport experiment. The total currents [calculated by
Eqs.~(\ref{electric-current})--(\ref{heat-current})] are thus a sum of
transport and magnetization parts,
\begin{equation}
\mathbf{j}^e = \mathbf{j}_{\text{tr}}^e + \mathbf{j}_{\text{mag}}^e \,
, \qquad \mathbf{j}^\sQ = \mathbf{j}^\sQ_{\text{tr}} +
\mathbf{j}^\sQ_{\text{mag}} \, .
\end{equation}
In the present case , following the arguments of
Ref.~\cite{Cooper-Halperin-Ruzin}, the magnetization electric and heat
currents which contribute to the total currents in
Eqs.~(\ref{jQy-over-Ex})--(\ref{jy-over-nablaTx}) are
\begin{equation} \label{mag-current}
\mathbf{j}_{\text{mag}}^e = c \frac{\partial \mathbf{M}}{\partial T}
\times (- \nabla T) \, , \qquad \mathbf{j}_{\text{mag}}^\sQ = c
\mathbf{M} \times \mathbf{E} \, ,
\end{equation}
where $\mathbf{M}$ is the equilibrium magnetization. The fluctuation
contribution to the magnetization is found either by
thermodynamics~\cite{reviews} or by calculating the current flowing in
the system in equilibrium in the presence of a varying system parameter
(such as the magnetic field or $T_c$).

By subtracting the magnetization currents, Eq.~(\ref{mag-current}), we
obtain the result for $\alpha_{xy}^\sSC$ (see Table~\ref{table:results}),
which diverges as the conductivity, $\alpha_{xy}^\sSC \propto \sigma_{xx}
^\sSC \propto (T - T_c)^{(d - 4) / 2}$. As expected, Onsager relations are
recovered by this subtraction. We note that the correction due to
magnetization currents is not small: In response to an electric field, the
magnetization current accounts for two thirds of the total heat current in
the weak magnetic field limit. In response to a temperature gradient, the
magnetization current is more singular than the transport current at $T_c$.

Finally, we consider the thermal conductivity tensor $\kappa$. The
longitudinal thermal conductivity $\kappa_{xx}^\sSC$ does not diverge
in any dimension, although it is singular. (In
Ref.~\cite{Varlamov-Livanov-90} a diverging result was obtained due to
the wrong form of the heat current vertex.) The transverse thermal
conductivity $\kappa_{xy}^\sSC$ has a logarithmic divergence at $T_c$
in two dimensions (similar to $\alpha_{xx}^\sSC$). In contrast with an
electric field, a temperature gradient does not give rise to a
magnetization heat current in the present case.

\emph{Cuprate data.}---We now compare our results for $\alpha_{xy}^\sSC$
with experiment. Consider the measurement of the Nernst effect: The sample
is placed in a temperature gradient $(- \nabla T)
\parallel \mathbf{\hat x}$ in the presence of a magnetic field $\mathbf{B}
\parallel \mathbf{\hat z}$. The transverse electric field
$E_y$ is measured in the absence of any transport electric current.
Imposing the condition $\mathbf{j}_{\text{tr}}^e = 0$ in
Eq.~(\ref{response-coefficients}) gives Nernst signal,
\begin{equation}\label{nernst}
  \nu = \frac{E_y}{(-\nabla T) B} =
  \frac{1}{B} \frac{\alpha_{xy} \sigma_{xx} - \alpha_{xx}
  \sigma_{xy}}{\sigma_{xx}^2 + \sigma_{xy}^2} \, .
\end{equation}
Taking the values for $\sigma$ and $\alpha$ from the Gaussian
contributions, we find that in this approximation the Nernst effect
tends to a constant at $T_c$, $\nu (T_c) = \alpha^\sSC_{xy} /
\sigma^\sSC_{xx} B$. However, for comparison with experiment the
fluctuation contributions, $\sigma^\sSC$ and $\alpha^\sSC$, should be
added to the normal state contribution, $\sigma^n$ and $\alpha^n$.
Moreover, we will consider temperatures which are not too close to
$T_c$, such that $\sigma^\sSC \ll \sigma^n \approx \sigma$.
Equation~(\ref{nernst}) may then be written in the low magnetic field
limit as
\begin{equation}\label{nernst-n}
  \nu \approx \nu^n + \frac{1}{B} \, \frac{\alpha_{xy}^\sSC}{\sigma_{xx}},
\end{equation}
The normal state Nernst effect $\nu^n$ is generally small due to a
cancellation between the two terms in Eq.~(\ref{nernst}).

\begin{figure}
\begin{center}
  \includegraphics[width=3in]{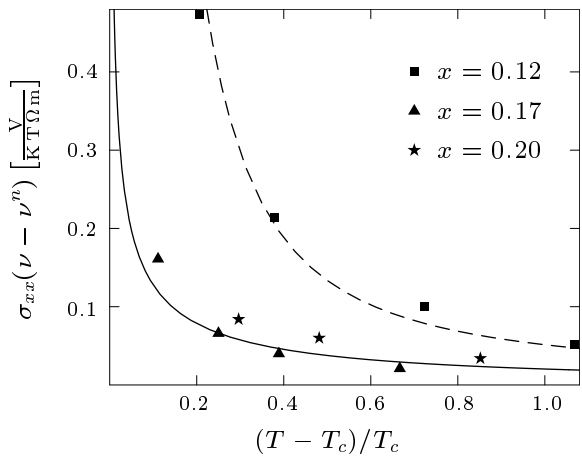}
\end{center}
\vspace{-0.3cm} \caption{\label{fig:data} Points are $\sigma_{xx} (\nu -
\nu^n)$ for different samples of La$_{2-x}$Sr$_x$CuO$_4$ \cite{Wang-Ong},
with $x = 0.12$ (underdoped, $T_c = 29$~K), $x = 0.17$ (near optimal
doping, $T_c = 36$~K), and $x = 0.2$ (overdoped, $T_c = 27$~K). Solid line
is theoretical value of $\alpha_{xy}^\sSC / B$, using $\xi_{ab}^{(0)} =
30$~\AA\ and an anisotropy of $\gamma = 20$. Dashed line is obtained using
a Hartree approximation (see text). \vspace{-0.3cm} }
\end{figure}

Eq.~(\ref{nernst-n}) is now used for a comparison with experiment in
the low magnetic field limit, presented in Fig.~\ref{fig:data} for data
measured on three samples of LSCO \cite{Wang-Ong}. The experimental
data shows $\sigma_{xx} (\nu - \nu^n)$, where $\nu$ and $\sigma_{xx}$
was measured at each temperature, while $\nu^n$ is deduced from an
extrapolation from the high temperature regime. As $\nu^n$ is very
small in this material, this extrapolation is expected to introduce a
relatively small error.

On the theoretical side, the result for $\alpha_{xy}^\sSC$ (see
Table~\ref{table:results}) is generalized for a layered superconductor
using the Lawrence-Doniach model,
\begin{equation}
\alpha_{xy}^\sSC = \frac{1}{6 \pi} \, \frac{e}{\hbar}
\frac{\xi_{ab}^2}{l_\sB^2 s} \, \frac{1}{\sqrt{1 + (2 \xi_c / s)^2}} \, .
\end{equation}
Here, $s$ is the interlayer spacing, and $\xi_{ab}$ ($\xi_c$) is the
coherence length in the direction parallel (perpendicular) to the
planes. Note that the only fitting parameters are the coherence length
$\xi_{ab}^{(0)}$ ($\xi_{ab} = \xi_{ab}^{(0)} \sqrt {T_c / (T - T_c)}
\,$) and the anisotropy $\gamma = \xi_{ab} / \xi_c$. In
Fig.~\ref{fig:data} $\alpha_{xy}^\sSC / B$ is plotted using
$\xi_{ab}^{(0)} = 30$~\AA\ and $\gamma = 20$ (cf., e.g.,
Ref.~\cite{Kimura-etal}).

As this comparison suggests, Gaussian superconducting fluctuations are
sufficient to explain the magnitude of the observed signal above $T_c$ in
the low magnetic field limit in the optimally doped and overdoped samples.
Two questions arise immediately: what is special about the cuprates and why
is the fluctuation contribution to the Nernst signal dominant at such high
temperatures? From our results, the answer to the the first appears to lie
in their anisotropy and in their smaller conductivity near $T_c$ which
together boost $\nu$ by two orders of magnitude from values we would
predict for bulk Al or Nb, despite their larger coherence lengths. In this
context it would be interesting to study low temperature superconductors
with high resistivities, such as the Nb films studied for fluctuation
effects by Hsu and Kapitulnik~\cite{Hsu-Kapitulnik}, where our formula
would indicate a Nernst signal comparable to optimally doped LSCO. In
answer to the second question we note that the Nernst signal is
particularly suitable for a fluctuation measurement due to the small
background signal in the normal state.

In the underdoped sample the measured signal is larger and cannot be
explained by Gaussian fluctuations if one uses the actual $T_c$. However,
this is the pseudogap region and in the interpretation considered in this
paper, the actual $T_c$ is suppressed from the mean-field transition
temperature $T_c^\sMF$ by non-Gaussian fluctuations and so a naive fit is
not justified; indeed, there is presumably a small effect of this kind even
at optimal doping which is unimportant in the region considered in our fit.

The increase in fluctuations with underdoping can be modelled by a growing
quartic term in the Ginzburg-Landau functional as well as by increased
two-dimensionality while the increase in $T_c^\sMF$ needs to be put in
directly. To get a sense of what these would do, we treat the resulting
problem via a self-consistent Hartree approximation (see, e.g.,
Ref.~\cite{Ullah-Dorsey}), which should be valid to some extent below
$T_c^\sMF$. This amounts to replacing $a$ by the self-consistent solution
of $\tilde a = a + b \langle | \psi |^2 \rangle$. That this is plausible is
demonstrated in Fig.~\ref{fig:data}, where we fit the data for the
underdoped sample (using a two-dimensional Hartree approximation, with
$T_c^\sMF=49$~K, $T_\sLambda=200$~K, $b / a_0 = 0.63 \hbar^2 / m^*$, and
$\xi_{ab}^{(0)} = 30$~\AA).

We thank Tom Lubensky and Steve Kivelson for valuable input, and Phuan
Ong and Yayu Wang for numerous enlightening discussions on their work.
We also thank the Rueff Wormser Fund, the Packard Foundation, and NSF
DMR-9978074, 9802468 and 0213706.

\end{document}